\documentclass[%
 reprint,
 amsmath,amssymb,
 aps,
prd,
]{revtex4-2}
\usepackage{graphicx}
\usepackage{dcolumn}
\usepackage{bm}

\usepackage{natbib}
\usepackage{enumerate}
\usepackage{comment}
\usepackage{amsfonts,amssymb,amsthm,array,amsmath,mathrsfs}
\usepackage{relsize}
\usepackage{cancel}
\usepackage{mathtools}
\usepackage[T1]{fontenc}
\usepackage{fancyhdr}

\usepackage{soul}

\usepackage[english]{babel}
\usepackage[utf8]{inputenc}

\usepackage{marginnote}
\usepackage[normalem]{ulem}

\usepackage{subfigure}
\newcommand{\sA}{\scriptscriptstyle\rm A}

\newcommand{\nn}{\nonumber}

\newcommand{\eq}[1]{(\ref{#1})}

\newcommand{\la}{\label}
\newcommand{\ba}{\begin{align}}
\newcommand{\ee}{\end{equation}}
\newcommand{\be}{\begin{equation}}

\def\12{\frac{1}{2}}
\newcommand{\p}{\partial}
\newcommand{\en}{\end{align}}
\newcommand{\e}{\epsilon}

\usepackage{color}

\def\green{\color{green}}
\def\blue{\color{blue}}

\begin{document}

\title{Hamilton Principle for Chiral Anomalies in Hydrodynamics  }
\author{P.B.~Wiegmann}
\affiliation{
Kadanoff Center for Theoretical Physics, University of Chicago,
5640 South Ellis Ave, Chicago, IL 60637, USA
}


\date{\today}





\begin{abstract}
We developed the spacetime-covariant  Hamilton principle for barotropic flows
of a perfect fluid in the external axial-vector potential conjugate to the
helicity current. Such flows carry helicity, a chiral imbalance, controlled
by the axial potential. The interest in such a setting is motivated by the
recent observation that the axial-current anomaly of quantum field
theories with Dirac fermions appears as a  kinematic property of classical hydrodynamics. Especially interesting effects occur under the simultaneous actions of the electromagnetic
field and the axial-vector potential. With the help of the  Hamilton principle,
we obtain the extension of the Euler equations by the axial potential and
derive anomalies in the divergence of the axial and vector current. Our approach provides a hydrodynamic expression for vector and axial currents and lays down a platform for studying
flows with a chiral imbalance and their anomalies.
 \end{abstract}
\maketitle
\section{Introduction}
Flows of an electrically charged barotropic perfect fluid feature two conserved
currents. One is the electric current whose spacetime components we denote
by \({J}^\alpha\) \begin{align}\la{1.01}
\p\!\cdot \!{J}=0.
\end{align}
Electric current is a conjugate to the electromagnetic vector potential \(A_\alpha\).
 The coupling with the electromagnetic field amounts to adding  the term
\begin{align}
\mathcal{S} ^{(0)}\to \mathcal{S} ^{(0)}  \,+\int_{\mathbb{R}^4} A\!\cdot\!
J\,\la{1.20}
\end{align} 
to the action of the perfect fluid $\mathcal{S}^{(0)}$. { The integral goes over spacetime $\mathbb{R}^4$ and we omit indices of contraction of spacetime vectors and covectors.}

{ Another conserved current is the helicity current \cite{carter}. In the neutral fluid helicity current is expressed through the fluid 4-momentum as } 
\begin{align}\la{1.1}
j_{\sA} ^\alpha=\e^{\alpha\beta\gamma\delta}p_\beta\p_\gamma
p_\delta,\quad \p\!\cdot\! j_{\sA}=0\,.
\end{align}    
Helicity current does not
depend on the spacetime metric and the expression \eq{1.1} holds for a relativistic  neutral
fluid as
well as for a Galilean fluid. For a relativistic fluid   \(p_\alpha\)
 is the conventional spacetime momentum.    We define the 4-momentum
for a Galilean fluid later in the text. 
The temporal component of the helicity current is familiar helicity density\begin{align}
j^0_{\sA}=\bm p\cdot \bm(\nabla\!\times\!\bm p),
\end{align}
where { \(\bm p\)} is the usual kinematic momentum (or, simply, momentum).

 The electric current is a spacetime vector and the helicity current is a
pseudovector  (or axial vector). Similar to the electric current one could
introduce a background  axial-vector potential \(A^{\sA}_\alpha\) conjugate
to the helicity current by adding the axial coupling to the
action \footnote{The convention for the sign is chosen in correspondence
to the equations of motions \eq{39} { and the definition of currents given by Eq. \eq{3.2}}}
\begin{align}
\mathcal{S} ^{(0)}\to \mathcal{S} ^{(0)}  \,-\int_{\mathbb{R}^4} A^{\sA}\!\cdot\!
j_{\sA}\,.
\end{align}
The axial-vector potential imposes a {\it chiral imbalance} by generating
flows with helicity.  Such flows become especially interesting in the electromagnetic
field background when the vector potential \(A\)  and the axial potential
\(A^{\sA}\) interfere. In this case, the action reads 
\begin{align}\la{6}
\mathcal{S} = \mathcal{S} ^{(0)}  \,+\int_{\mathbb{R}^4} [A\!\cdot\!
J- A^{\sA}\!\cdot\!
j_{\sA}]\,.
\end{align}
 
This is the setting we study in this paper.
The paper aims  to shed  some light on the hydrodynamic meaning of the  important kinematic phenomena  commonly referred
to as {\it chiral anomalies}. 

The chiral anomalies are fundamental phenomena
of quantum field theory with fermions, such as QED. In  the
last decade it had been important developments in a hydrodynamic description
 of relativistic   quantum field theories of chiral (Weyl) fermions
 consistent with chiral anomalies  (see, \cite{son2009hydrodynamics,stephanov2012chiral,jensen2013thermodynamics,haehl2014effective}
and references therein). { Recently} in \cite{abanov2022axial,wiegmann2022chiral} it was found that { the} chiral anomaly  { of} Dirac fermions is a property of the Euler equation for barotropic fluid.

Because the helicity current \eq{1.1} involves derivatives of the momentum
the axial coupling changes the traditional  relations between
electric current \(J\),  the momentum \(p\), and the velocity held in a perfect
fluid.  Say, in a non-relativistic perfect fluid    \(\bm p=m\bm v\) and
  \(\bm  J=en\bm v=(en/m)\bm p\). This relations will be deformed by  background
potentials \(A_\alpha\) and \(A^{\sA}_\alpha\) (see, \eq{3.40} below).  Furthermore,
and most importantly,  the electromagnetic potential in turn gives { a} feedback
to the axial current \eq{1.1}. In Ref. \cite{abanov2022axial} it was shown
that  the axial current  is modified by the electromagnetic field as \begin{align}\la{1.6}
j_{\sA} ^\alpha=\e^{\alpha\beta\gamma\delta}p_\beta(\p_\gamma
p_\delta+F_{\gamma\delta})\,.
\end{align} 
The temporal component of the axial current now  reads\begin{align}
j^0_{\sA}=\bm p\cdot \bm(\nabla\!\times\!\bm p+2\bm B)\,,
\end{align} where \(\bm B\) is the magnetic field and \(F_{\alpha\beta}=\p_\alpha
A_\beta-\p_\beta
A_\alpha\) is the electromagnetic field tensor. 

{ Now, in the electromagnetic background, } the axial current is no longer divergence-free. In \cite{abanov2022axial}
it was shown that although the { axial} current is not conserved its divergence does
not depend on dynamic fields. Rather, it is { solely}  controlled by the electromagnetic
field as
\begin{align}\fbox{$
\p\!\cdot\!j_{\sA}=    \tfrac  12{}^\star\!F^{\alpha\beta}\!\,\!F_{\beta\alpha}$}
 \,\la{-1}
\end{align}   
or \begin{align}
\p\!\cdot\!j_{\sA}=2\bm E\!\cdot\!\bm B
\end{align}where \(\bm E\) is the electric field,   and \(^\star\! F^{\alpha\beta}=\tfrac
12 \e^{\alpha\beta\delta\gamma}F_{\delta\gamma}\)
is the dual field tensor.

The formula \eq{-1}  is identical to the celebrated {\it axial-current anomaly} of QED.
It appears that     among two independent gauge symmetries of the classical Lagrangian of QED \begin{align}\la{1.8}
&A_\alpha\to A_\alpha+\p_\alpha\varphi\,,\\
& A_\alpha^{\sA}\to A_\alpha^{\sA}+\p_\alpha\varphi^{\sA}\la{1.9}
\end{align} 
only the vector gauge transformation  \eq{1.8}, gives rise to
conserved current, that is the electric current \eq{1.01}. { The axial gauge invariance  \eq{1.9} does not.} In this paper, we describe a similar situation in hydrodynamics. { We will see that the}  Euler equation deformed by the axial coupling  is invariant under both transformations, however, solutions of the equation which describe physical relevant flows are not.
The conflict between the vector and the axial gauge symmetries  in quantum field theory with Dirac fermions had been
discovered in  1969 by Adler
\cite{adler1969axial}
and Bell and Jackiw \cite{bell1969pcac}.

We recall
 that the Dirac
multiplet consists of components of the left and right-handed chiral particles.
The quantum states of a massless Dirac theory  are  characterized by the
vector current \(j\) and  the axial current
\(j_{\sA}\), the sum and the difference of the currents of left and
right components.  Both currents are conserved and could be coupled with
a vector and an axial-vector potential. 
However,  due to subtle quantum effects
 the electromagnetic background field prevents the conservation of the axial
current.  
 Adler and Bell and Jackiw found that in units of the flux quantum
\(\Phi_0=he/c\), where  \(h\) is the Planck constant the divergence of the axial
current
 is given precisely by the Eq.\eq{-1}.
Furthermore, this result is largely insensitive to interaction and holds
with or without axial potential.
 
We encounter a remarkable correspondence between the kinematic properties
of the classical fluid and the properties of quantum states
of  Dirac fermions. This must not come as a surprise. It is broadly known
that in one spatial dimension the Hilbert space of the Dirac fermions
is identified with that of a Bose scalar field, whose dynamic is equivalent to { that of } a
compressible fluid.  We see that the part of this correspondence that is based on geometry extends
to higher even
spacetime dimensions.  

The kinematic properties of hydrodynamics are closely related to the geometry
of flows \cite{arnold2008topological}.  Likewise, anomalies of quantum field
theory, too are related to the geometry of quantum states. Both are largely decoupled
from the dynamics.   An apparent
relation between the geometric properties of classical flows and
quantum states { revealed}  by the anomalies is noteworthy.  In this paper, we further
develop this subject. Our motivation comes from the fermionic quantum field theory, but the paper is written from the side of fluid mechanics with a minimal
appeal to the vast literature on anomalies in quantum field theory. 

Part of the results of this paper appeared in  \cite{wiegmann2022chiral},
where we discussed the  coupling with the time-independent
axial chemical potential
\(\mu^{\sA}=A^{\sA}_0\). Inclusion the  spatial components  \(\bm A^{\sA}\)
of the axial potential requires a spacetime covariant approach we developed here. 

 We use the  Hamilton principle on the Lie group of spacetime diffeomorphisms \({\rm Diff}(\mathbb{R}^4)\). With the help of the Hamilton principle, we obtain the spacetime-covariant equations
of motion,  the conservation laws, and the hydrodynamic representation of various currents by Eulerian fields. A brief account of essential results presented
here was reported in \cite{abanov2022anomalies}, where we used the spacetime-covariant Hamilton principle in the dual space.  See, also a related paper \cite{monteiro2015hydrodynamics}. For
references, we collect the major
formulas in the last section of the paper.

The paper is organized as follows. In the first few sections, we assume that
the axial anomaly equation \eq{-1} holds and on this basis draw general consequences.  Then we describe the class of
fluid systems which feature the axial-current anomaly.  Such systems are
referred to as {\it uniformly canonical}.
The simplest system in this class is perfect barotropic fluid and it remains uniformly canonical under the axial
coupling. We describe the Hamiltion principle for this class of fluids
and with its help obtain the equations of motion.

The notion of uniformly canonical fluid systems was introduced by Carter
 \cite{carter}. These fluids are Hamiltonian systems, that do not possess any advected local { scalar} fields, such as, e.g., entropy. { In this case, the four components of the spacetime canonical momentum
 \begin{equation}
 \pi_\alpha= p_\alpha+
A_\alpha\,.\label{1.10}
\end{equation}
 are the only dynamical fields. The configuration space of such systems is identified with the group of spacetime diffeomorphisms  \({\rm Diff}(\mathbb{R}^4)\) 
and its dynamics follow from the Hamilton principle with the action defined on the Lie algebra of vector field
  (flow field) \({\rm Diff}(\mathbb{R}^4)\) or in its dual space of the canonical momentum.    }

 \section{Perfect barotropic fluid} 
This section aims to establish notations and to introduce the 4- momentum for a Galilean fluid.

  Flows of a  perfect charged  fluid
in the electromagnetic field  are governed by the Euler equation and the
continuity
equation\begin{align}\la{0}
\begin{cases}&\dot{\bm v }+(\bm v\!\cdot\!\bm \nabla)\bm v+\rho^{-1}\bm\nabla P=\bm
E+\bm v\times \bm B\mathrm\,,\\
&\dot n+\bm\nabla\bm n=0\,,
\end{cases}
\end{align} where \(\rho=mn\) is the mass density, \(n\) is the particle
number density,  \(\bm v\) is the velocity of the fluid, \(P\) is pressure
and we denote 
    \(n^\alpha=(n,n\bm v) \)
  the particle number 4-current. 
The same equations   hold  in  a neutral fluid
rotating with the Larmor frequency \(\bm
B/m\) placed in the potential \(A_0\).
A barotropic fluid is singled out by the condition that pressure is a function
of the particle number $n$. Then \(\bm\nabla
P\) and \(\bm \nabla n\) are collinear (this  property is the origin
of the term barotropic) and
 \(\bm\nabla P/\rho=\bm\nabla\mu\), where
\(\mu=\p\varepsilon/\p
n\) is the chemical potential  and  \(\varepsilon\)
is the fluid internal energy.  

 We will use the spacetime covariant formalism,
based on the notion of the spacetime fluid 4-momentum. In the case of the relativistic
fluid, this is a usual 4-momentum. It is a spacetime covector constrained
by the condition \(p_\alpha p^\alpha=-(mc+ c^{-1}\mu)^2\).    In the non-relativistic
(Galilean) fluid the notion of the  4-momentum  may be less familiar. It could be obtained
as a
non-relativistic limit of the above formula and dropping \(mc^2\). We find that \(p_0\) is (minus) Bernoulli function{, that is the energy per particle } 
\be p_\alpha=(p_0,\bm p):\quad p_0=-\mu-\bm p^2/2m,\quad \bm p=m\bm v\,.\la{1.11}\ee

Alternatively, \(p_0\)  could be introduced as a canonical conjugate to the
particle number \(n\). Consider the action of a Galilean  perfect fluid
 and treat it as a function of  particle number  4-current  \be n^\alpha=(n,n\bm
v)\la{2.301}\,.\ee
The action reads
\be \mathcal{S}^{(0)}=\int_{\mathbb{R}^4}\left(\tfrac
m{2n}\bm n^2-\varepsilon[n]\right).\ee 
Then the 4-momentum is defined through
the variation \begin{align}
\mathcal{\delta S} ^{(0)}=\int_{\mathbb{R}^4}p_\alpha\delta n^\alpha\,.
\end{align} 
Because the axial
current \eq{1.6} depends only on the momentum, the axial coupling does not
affect  \eq{1.11}. 

The Legendre transform on the action defines the spacetime version of the
``Hamiltonian''  \(H^{(0)}=\int_{\mathbb{R}^4}(p\!\cdot\!
n)-\mathcal{S}^{(0)}\).  That gives  \begin{align}\la{2.6}
H^{(0)}=-\int_{\mathbb{R}^4} P \,,
\end{align} 
where \(P=n\mu -\varepsilon\) is the pressure treated as a function of the 4-momentum under the relation \eq{1.11}. { The integral of the  pressure  \eq{2.6} treated as a functional of  4-momentum was used as a variational functional for the perfect fluid in \cite{schutz1970perfect,carter1988standard}. }

An equivalent form of equations of motion  \eq{0}  is the conservation laws
of energy and  momentum  
 \begin{align}\p_\alpha T_\beta^\alpha=F_{\beta\gamma}J^\gamma\,,
\la{1.500}
\end{align}
where \(T_\alpha^\beta\) is the momentum-stress-energy
tensor (or, simply, stress tensor)  and \(J^\alpha\)  is { conserved}  electric current.
The r.h.s. in \eq{1.500} is the Lorentz force. In perfect fluid (in units of the electric charge) the electric current equals the particle
number current
\(J^\alpha=en^\alpha\). { Then, } the  stress
tensor of the perfect fluid is
\be T_\alpha^\beta= 
n^\beta  p_\alpha+P\delta_{\alpha}^\beta\,.\la{1.100}\ee The formulas (\ref{1.500},\ref{1.100})
hold in the  relativistic
case  with the particle number $n$ replaced by  \(n/\sqrt{1-\bm
v^2/c^2}\). In the relativistic case, the stress tensor is symmetric. 

    Four equations
\eq{1.500}
fully describe the barotropic fluid as the continuity equation  { in \eq{0}}  follows from
\eq{1.500}.
This would not be the case for a baroclinic fluid.   

{ In the rest of the paper, we seek a deformation of the two equivalent forms of equations of motions \eq{0} and  \eq{1.500} by the axial vector potential $A^{\sA}$.}

\bigskip

  \section{Symmetries of equations of motion}
According to basic principles, equations of motion and their solutions are invariant under the gauge transformations \eq{1.8} and the spacetime diffeomorphisms. The axial gauge transformation (\ref{1.9})
 is more subtle. We will see that equations of motion are locally expressed through  the Eulerian  { fields (the 4-momentum $p^\alpha$)} and the field tensors 
 \(F_{\alpha\beta}\) and 
 \(F_{\alpha\beta}^{\sA}\)\,.  Then at given { $p^\alpha$}, the equations of motion do not change under the axial gauge transformations (\ref{1.9}).  However, as a consequence of the anomaly solutions of the equations, the flows, could be different at gauge equivalent values of the axial potential \(A^{\sA}\) and \(A^{\sA}+\p\varphi^{\sA}\). From this perspective, the axial-gauge transformations are not true symmetry. 

 We assume that under the action of the axial transformations the Lagrangian density changes only by a full spacetime derivative and explore the consequences. Later we justify this assumption. 
\subsection{Vector current and chiral anomaly }\la{V}
Under the axial gauge transformation (12) { and due to the axial-current anomaly  \eq{-1} the axial coupling taken at a fixed momentum}  transforms as
  \begin{align}\nn
\delta^{\sA} \Big(\int_{\mathbb{R}^4}&A^{\sA}\cdot
j_{\sA}\Big) { =}\int_{\mathbb{R} ^4}\p\varphi^{\sA}\cdot
j_{\sA}=-\int_{\mathbb{R} ^4}\varphi^{\sA}(\p\!\cdot\!
j_{\sA})=\\
&- \tfrac12\int_{\mathbb{R} ^4}\varphi^{\sA} F\!\cdot\!{}^\star\!F=2\int_{\mathbb{R}
^4}A_{\alpha} \,{}^\star
\! F^{\alpha\beta}\p_\beta\varphi^{\sA}\,. \la{23}
\end{align}
  If we demand that the action (up to boundary terms) is invariant { under \eq{1.9}}, then \eq{23}   must  be compensated  by the transformation of the electric
current 
\begin{align}\la{3.21}
\delta^{\sA} J^\alpha{  |_p}=2\,{}^\star
\! F^{\alpha\beta}\p_\beta\varphi^{\sA}.
\end{align}
We see that as a result of the anomaly the electric current, an observable quantity, is not an axial-gauge invariant.  
We introduce the axial-gauge invariant   \emph{vector
current} \begin{align}
 \la{2.300}
         j^\alpha:=J^\alpha-2\, {}^\star\!F^{\alpha\beta}A_\beta^{\sA}\,.
\end{align}   
   Because
the electric current  conserves, the vector current does  not 
\begin{align}\la{-2}\fbox{$
\p\!\cdot\! j = F\cdot {}^\star\! F_{\sA}$} \end{align}

Eq.\eq{-2} is the consequence of the axial-current anomaly { .}   In this context the anomaly says that  the conserved electric current { $J$}  is not axial-gauge invariant and  the axial-gauge invariant vector current { $j$}  is not conserved. Together Eqs.(\ref{-1},\ref{-2}) sometime referred to as {\it chiral anomalies}. 

Both currents \(J\) and \(j\) are physical observables. The difference between
them depends only on the 
 background fields.  It stays the same for different
flows occurring at the same external fields.   This property suggests that
under the axial coupling, the fluid is necessarily connected with a reservoir,
  a spectator medium whose motion is solely determined by the background fields.
Then, the  term \(2\, {}^\star\!F^{\alpha\beta}A_\beta^{\sA}\) in \eq{2.300} should be associated with  the electric
current run through the
reservoir and the vector
current \(j\) { is associated }     with  the  
electric current conducted by the fluid. 
Since the fluid and the reservoir are open systems their currents taken independently
do not conserve. The total electric current does.

We illustrate this  interpretation by the experimentally established
 Landauer-Sharvin effect  of electronic ballistic transport
(see, \cite{alekseev1998universality,frohlich2000new}). In the  Landauer-Sharvin setting, the electric current ballistically runs through a wire whose open ends are
connected to metallic leads (the reservoir).  Then the  one-dimensional version of the formula  \eq{-2}  \(\p\cdot  j
=2E\), written in the static case as \(j=2U\),  where \(U\) is the voltage drop between the leads {($E=\nabla U$)} describes the universal conductance. Restoring the units the universal conductance reads  \(2e^2/h\).

\subsection{Spacetime diffeomorphisms:  equations of motions in
the form of conservation laws } 
Next, we draw the consequences of the invariance under spacetime diffeomorphisms.
 On a general ground, a system coupled to vector potentials features the relation between the stress tensor { of the perfect fluid \eq{1.100}} and external potentials
 \begin{align}
 \la{2.1}
        \p_\alpha {}T^\alpha_\beta = F_{\beta\alpha}J^\alpha- A_\beta\, (\p\!\cdot\!
J)+F^{\sA}_{\beta\alpha}j_A^\alpha- A_\beta ^{\sA}\,(\p\!\cdot\! j_{\sA})\,.
\end{align}
At the end of this section, we recall the derivation of \eq{2.1}. 

If the currents are conserved the terms with divergence drop and the r.h.s. of \eq{2.1}
is a sum of two Lorentz forces.
In our case the electric current is conserved but the axial current is not.
Given the axial-current anomaly \eq{-1}  and with the help of the identity
valid for any two antisymmetric tensors 
\begin{align}\la{3.121}
2(^\star m^{\beta\gamma}l_{\gamma\alpha}+^\star\! l^{\beta\gamma}m_{\gamma\alpha})={ \delta_\alpha^\beta}({}^\star
m^{\delta\gamma}l_{\gamma\delta})
\end{align}
 we write the last term in \eq{2.1} as \(-A_\beta ^{\sA}{ \,\p\!\cdot\! j_{\sA}= -
   \tfrac  12A_\beta ^{\sA}\, F\!\cdot\!{}^\star F}=-2F_{\beta\alpha}
\,{}^\star\!
F^{\alpha\gamma}A_\gamma ^{\sA}\) [we applied the identity { \eq{3.121}}  by setting both
tensors \(m,l\) to be equal to the field tensor \(F\). See Appendix for the origin of this identity].  This term joins the first
term in \eq{2.1} bringing the equation to the form \begin{align}
 \la{39}
        \p_\alpha T^\alpha_\beta = F_{\beta\alpha}j^\alpha+F^{\sA}_{\beta\alpha}j_A^\alpha\,,
\end{align} 
where \(j\) is the axial-gauge invariant  vector current introduced in Sec.\ref{V}, Eq.\eq{2.300}.
We observe that  this equation is expressed through the axial field tensor $F^{\sA}$ { and the 4-momentum}. This justifies the assumption made at the beginning of the section.
   
The axial-current anomaly gives a different meaning to the
 Lorentz force as the currents entered the Lorentz force in 
 \eq{39} do not conserve. Their divergence is determined by the chiral anomaly (\ref{-1},\ref{-2}).
 
These equations are valid for the fermionic quantum field theory and the classical fluid alike. In the case of the fluid we know the stress
tensor  (see, Eq.\eq{1.100}) and the axial current \eq{1.6}.    This gives us a set of
six equations for four components of the vector current which remains undetermined. It is quite remarkable that the over-complete set is consistent. An observation that the conservation laws and an imposed anomaly completely determine the currents was first made by Son
and Surowka in Ref.\cite{son2009hydrodynamics} for a related
problem. We do not pursue this avenue. In the next section, we obtain the vector current differently {( see, Sec.\ref{R}).}
{ Also, it remains to be seen that Eq.\eq{39} follows from the action \eq{6}. We will do this in the Sec. \ref{B}.
}

Now we recall the standard arguments leading to Eq.\eq{2.1}.
The equation follows from   the invariance  of  the action \(\mathcal{S}\)  under the group of spacetime diffeomorphisms \({\rm
Diff}(\mathbb{R}^4).\) The \(\mathcal{S}^{(0)}\) part of the action depends on the spacetime metric, and the coupling depends on the external potentials. External potentials are spacetime covectors. This is obvious for the vector potential \(A_\alpha\).  It is also true for the axial potential because the components of the axial current constitute the 3-form \(j_A=p\wedge d( p+2A)\). Then the coupling   \(\int_{\mathbb{R}^4}
A ^{\sA}\!\cdot\!j_{\sA}\) be a spacetime scalar if \(A^{\sA}_\alpha dx^\alpha\) be 1-form and \(A^{\sA}_\alpha \) is a spacetime covector.

Under a diffeomorphisms \(x^\alpha\to x^\alpha+\e^\alpha(x)\) the transformation
of the  spacetime, covectors are given by the action of
the Lie derivative along the vector field with components 
\(\e^\alpha\). They are 
\footnote{Transformations of cotensor or differential forms under the diffeomorphisms is given by the Cartan formula  \eq{2.8} in Appendix.}
 \begin{align} 
 \delta_{ \e} A_\alpha^{\sA}&=\e^\beta F_{\beta\alpha}^{\sA}+ \p_\alpha(\e\!\cdot \!A^{\sA})\,,
 \la{34}\\
        \delta_{ \e} A_\alpha&=  \e^\beta F_{\beta\alpha}+ \p_\alpha(\e\!\cdot \!A)\,.\la{35}
\end{align} 
 Then the transformation of the action under combined transformations of the external potentials,   the spacetime frame, and dynamical fields vanishes. The contribution of the transformations of the dynamical fields will be dropped if we evaluate the result on the equations of motion (the Hamilton principle). 

 Given that { under coordinate transformation at fixed momentum} \(\delta_{ \e}\mathcal{S}^{(0)}=\int_{\mathbb{R}^4}T^\beta_\alpha \p_\beta\e^\alpha\), we { have 
 \begin{align}
{ \delta_\e\mathcal{S}}=\int_{\mathbb{R}^4}[T^\beta_\alpha \p_\beta\e^\alpha+J\!\cdot\!\delta_{ \e}
A+j_{\sA}\!\cdot\delta_{ \e} A ^{\sA}]\,{ .}\la{341}
\end{align}
Reporting
(\ref{34}-\ref{35}) in equation \eq{341} { and setting $\delta_\e\mathcal{S}|_{\text{EOM}}=0$ we obtain}   \begin{align}
0=\!\int_{\mathbb{R}^4}\e^\alpha[-\p_\beta T^\beta_\alpha+F_{\beta\alpha}J^\alpha\!-\!
A_\beta \,\p\!\cdot\!
J\!+\!F^{\sA}_{\beta\alpha}j_A^\alpha-\! A_\beta ^{\sA\,}\p\!\cdot\!j_{\sA}]\,\nn
\end{align}
and \eq{2.1}. 
\subsection{ Energy flow driven by electric current}
This section aims to clarify the relationship between the electric
current \(J\) and the vector current \(j\). 

 Let us assume that the axial potential is time-independent. Then  we may
 speak about the energy of the  combined  system, the
reservoir and the fluid. The energy density and the energy flux  of such
system are \begin{align} {\mathcal{E}} =\mathcal{}e-A_0^{\sA}j^0_{\sA},\quad
{\bm J}_\mathcal{E}=\bm
j_e-A_0^{\sA}\bm j_{\sA}\,.\la{37}
\end{align}   where  \(e=n\tfrac{\bm p^2}{2m}+\varepsilon,\,\bm
j_e=n\bm v(\tfrac{\bm p^2}{2m}+\mu) \) are the energy density and
the energy flux of the perfect
fluid. The temporal component of Eq.\eq{39}
is  the equation for the  fluid energy flux \begin{align}\la{5.9}
\dot e+\bm\nabla\bm
j_e=\bm E\cdot\bm j+\bm E^{\sA}\cdot\bm j^{\sA}.
\end{align}
 We may rewrite this equation in terms of the energy flux \eq{37} of the combined system \begin{align}
\dot{{\mathcal{E}}}+\bm\nabla\!\cdot\!{\bm J}_\mathcal{E}=\bm
E\!\cdot\!\bm J\,.
\end{align}
 We see that the flux is driven by the electric current    \(\bm J\). This suggests identifying \(\bm J\)
  with the electric current of the combined system. 
  
\section{Hydrodynamic representation of the currents}  
In this section, we find the explicit expression
for the electric current in terms of Eulerian { fields} . The reciprocal consistency relation offers an economic approach.

\subsection{Reciprocal relation}\la{R}
 
    The reciprocal relation states
that the cross-variations of
the currents evaluated on the equations of motion
are equal. That is
\begin{align}
\quad\int_{\mathbb{R} ^4}(\delta J_{\rm }^\alpha)\delta A_\alpha=\int_{\mathbb{R}
^4}(\delta j_{\sA}^\alpha)\delta
A_\alpha^{\sA}.
\end{align}
Here the variation of \(J\) is taken at a fixed \(A\) and the variation of
\(j_{\sA}\) is taken at a fixed \(A^{\sA}\) and both variations are taken on  the equations
of motion (EOM).

The reciprocal principle follows from the Hamilton principle and the definitions
of currents as the first variation of the action evaluated on the equations
of motion { (EOM) }\footnote{Different signs in \eq{3.2} and \eq{6} reflect the
contribution of \(\mathcal{S}^{(0)}\), Sec.VIII B.}\begin{align}
\quad\delta\mathcal{S}|_{\rm EOM}=\int_{\mathbb{R}^4}[J_{\rm }\!\cdot\!\delta A+j_{\sA}\!\cdot\!\delta
{ A^{\sA}}]\,.\la{3.2}
\end{align}
Now, given the explicit form of the axial current  \eq{1.6} the
reciprocal relation determines the electric current. Given \eq{1.6} we vary
it over \(A\) at a fixed canonical momentum \(\pi\).  [The
variation at a fixed canonical momentum \(\pi\) sets the
variation on EOM].  We obtain  (with the help of {\blue \eq{3.21}}) )
\begin{align}\la{3.3}
\delta^{ \sA} J^\alpha|_\pi=\delta {}^\star \! F_{\sA}^{\alpha\beta} p_\beta
        +2{}^\star \! F^{\alpha\beta}\delta A^{\sA}_\beta \,.
\end{align}
 Now we can integrate \eq{3.3} with a condition that in a perfect fluid, the electric current is equal to the
particle number current. This gives \begin{align}\la{3.40}
&J^\alpha=n^\alpha+{}^\star \! F_{\sA}^{\alpha\beta} p_\beta
        +2\,{}^\star \! F^{\alpha\beta} A^{\sA}_\beta\,,\\
        &j^\alpha=n^\alpha{+}{}^\star \! F_{\sA}^{\alpha\beta}p_\beta\,.\la{3.50}
\end{align}   
  
These are the hydrodynamic representation of the electric
current (and the vector current). Given that, we obtain the
explicit form of the fluid action { \eq{6}}. We write it in terms of the
vector and the axial currents
\begin{align}\la{43}
\mathcal{S} = \mathcal{S} ^{(0)}  +\int_{\mathbb{R}^4} \left(A\!\cdot \!j-A^{\sA}\!\cdot\!
j_{\sA}\right)+\int_{\mathbb{R}^4}2A_\alpha{}^\star \! F^{\alpha\beta} A^{\sA}_\beta\,,
\end{align}
and 
in terms of the Eulerian fields
\begin{align}\la{3.10}
\mathcal{S} = \mathcal{S} ^{(0)}+\int_{\mathbb{R}^4} \Big(A_\alpha n^\alpha-A_\alpha^{\sA}
\e^{\alpha\beta\gamma\delta}p_\beta\p_\gamma p_\delta\Big)+&\\
\int_{\mathbb{R}^4}-\left(A_\alpha{}^\star \! F_{\sA}^{\alpha\beta}+2A_\alpha^{\sA}\,{}^\star\!
F^{\alpha\beta} \right) p_\beta
        +\nn&\\
        \int_{\mathbb{R}^4}2A_\alpha{}^\star \! F^{\alpha\beta} A^{\sA}_\beta\,.&\nn
\end{align} 
We wrote the action in the form which separates orders of the external field.
 
\section{Uniformly canonical fluid systems}
  A barotropic perfect fluid is the simplest example of a  general class
of fluids referred to as {\it uniformly
canonical} \cite{carter}.  These fluids do not possess advected scalar field (a Lagrangian scalar),
such as entropy, or if there is one,  it must be chosen to be uniformly the same on all streamlines (if
the Lagrangian scalar is entropy such flows are called homentropic).   In
this case, the number of independent  Eulerian fields equals the dimension
of spacetime, and the 4-momentum could be chosen to characterize the flow. The barotropic fluid under axial coupling belongs to this class. 

Equivalently,  the uniformly canonical systems   could  be defined by identifying their configuration space with the group of space-time diffeomorphisms    \({\rm Diff}(\mathbb{R}^4)\) and its Eulerian  { fields}  (currents) with the Lie algebra   
 \({\rm Diff}(\mathbb{R}^4)\), or the dual to the Lie algebra (the momentum).
  
Uniformly canonical systems possess remarkable geometric properties our analysis
 relays upon. Vorticity
surfaces of such flows are integral and form a foliation of spacetime. A
major consequence of this property is the conservation of helicity current
\cite{carter} which we discussed in the next section (also, see Appendix).

 Let us start with the perfect barotropic fluid. We may write the 
 Euler equation \eq{0}  in the  form \begin{align}
   n( \dot{\bm\pi}-\bm\nabla\pi_0)-\bm n\times(\bm \nabla\times\bm\pi)\,=0\,.
 \la{29}
\end{align}        
This form of  the 
 Euler equation emphasizes a different role of the particle
number current \( n^\alpha\) \eq{2.301} and the   canonical momentum \eq{1.10}.
{ A corollary of this equation is 
\be\la{291}
\bm n\cdot ( \dot{\bm\pi}-\bm\nabla\pi_0)=0\,.
\ee
}
 Introducing  canonical vorticity tensor
\be \Omega_{\alpha\beta}=\p_\alpha \pi_\beta-\p_\beta \pi_\alpha\la{4.200}\ee
we write Eq.\eq{29} { and its corollary \eq{291} }in a remarkably compact form   
 \begin{align}\la{1.21}
\mathcal{J}^\alpha \Omega_{\alpha\beta}=0\,, 
\end{align}
where we renamed the particle number current \(n^\alpha\) by \(\mathcal{J}^\alpha\). 
 This  is  the Carter-Lichnerowicz
equation (see, {  \cite{carter1988standard}} and \cite{rieutord2006introduction,markakis} for a review). It must be complimented by the continuity equation \begin{align}\la{4.4}
\p\!\cdot\!\mathcal{J}=0\,.
\end{align}

 Eq.\eq{1.21}   with a not specified  current \(\mathcal{J}\) used as a definition of  uniformly canonical fluid systems by Carter \cite{carter}. Additional
information
about the relation between the conserved current $\mathcal{J}$ and the momentum
defines a specific { fluid} system. This relation involves the spacetime metric, while
the equation  \eq{1.21} does not. In the perfect fluid \(\mathcal{J}^\alpha=n^\alpha\)
but, generally, it could be different.

An easy consequence of the Carter-Lichnerowicz
the equation is the Helmholtz law: vorticity \(\Omega\) is advected along the vector field defined by \(\mathcal{J}\)  (see, \cite{markakis} and the Appendix). This property gives the meaning to \(\mathcal{J}\): a  vector field associated with  \(\mathcal{J}\) defines the streamlines.
We, therefore, refer to  \(\mathcal{J}\)  as a {\it flow  field}, { or as a contravariant vector flow field.}

An example of a  non-uniformly canonical system is a baroclinic fluid. In
this case, the r.h.s of \eq{1.21}  equals \(n(\p\varepsilon/\p S)\p_\beta
S\), where \(S\) is a Lagrangian scalar,  such as the entropy per particle.  
 
\section{Spacetime covariant Hamilton  principle for  uniformly canonical
fluid systems}
Now we turn to the Hamilton principle which leads to the Eq.\eq{1.21}. We consider the action
 \(\mathcal{S}[\mathcal{J}]\) as a functional of the flow field \(\mathcal{J}\)
 and define the canonical momentum  as a conjugate \begin{align}\la{49}
\delta\mathcal{S}=\int_{\mathbb{R}^4}\pi\!\cdot\!\delta \mathcal{J}\,.
\end{align}
{ in the dual space,} cotangent bundle of \(\mathbb{R}^4\). Often it is more convenient to operate in a dual space considering a spacetime covariant
Hamiltonian, the extension of the functional \eq{2.6}. It is a functional of \(\pi\) 
defined by the Legendre transform\begin{align}\la{50}
H[\pi]=\int_{\mathbb{R}^4}\pi\!\cdot\!\mathcal{J}-\mathcal{S}\,.
\end{align}
  The equations of motions follow
from the  Hamilton principle
\begin{align}
&\delta\mathcal{S}[\mathcal{J}]=0\,,\la{5.61}
\end{align}
Defined  in the tangent
bundle of \(\mathbb{R} ^4\), or equivalently from the { Hamilton} principle  defined  in the cotangent
bundle of \(\mathbb{R} ^4\)
\begin{align} \delta H[\pi]=0\,.\la{5.71}
\end{align}

 \subsection{Admissible variations and the Lie algebra of spacetime diffeomorphisms}\la{41}
 In 1966 Arnold \cite{arnold} developed a framework in which a set of possible fluid motions (a configuration space) { is identified} as an infinite-dimensional Lie group of diffeomorphisms. Within this  framework, { the} configuration space  of uniformly canonical systems  is the group manifold of  spacetime
diffeomorphisms
   \({\rm Diff}(\mathbb{R}^4)\).   The elements of the Lie algebra  
   \({\rm Diff}(\mathbb{R}^4)\) are vector
fields { associated with the flow field  \(\mathcal{J}^\alpha \sqrt{|g|}\), where \( |g|\) is the determinant of the spacetime metric.  }   

  The Lie algebra \({\rm
Diff}(\mathbb{R}^4)\) is endowed with the Lie bracket, or the commutator of vector fields. If  \(X(x)=X^\alpha
(x)\p_\alpha\) and \(Y(x)=Y^\alpha
(x)\p_\alpha\)  are two vector fields then their commutator is
  \begin{align}
[X,\, Y]:=\left( X^\alpha\p_\alpha Y^\beta-
Y^\alpha\p_\alpha X^\beta\right)\p_\beta\,.
\end{align}
 
 The Lie bracket defines the transformation of the vector field under spacetime diffeomorphism. { Denote} a spacetime diffeomorphism
  \be  x^\alpha\to  x^\alpha+
\e^\alpha( x).\,\la{59}\ee 
 Then the     change of \(X\)   along the flow generated  by the vector
field \(\e=\e^\alpha\p_\alpha\) known as the Lie derivative $\mathcal{L}_\e X=\delta_\e X$, is \begin{align}
\delta_\e X\,=[\e,X]\,.\la{4.40}
\end{align} 
Explicitly  \begin{align}
\delta_\e X^\alpha=(\e\cdot\p)\,X^\alpha-(X\!\cdot\p)\,\e^\alpha
\,.\la{5.300}
\end{align}
Taking into account the transformation  element as \(\delta(\sqrt{|g|})|_{|g|=1}=\p\cdot\e\) we obtain the  transformation of the current \begin{align}
\delta\,\mathcal{J}^\alpha=\p_\beta(\e^\beta\mathcal{J}^\alpha-\e^\alpha\mathcal{J}^\beta)+\e^\alpha\mathcal{\p_\beta J}^\beta
\,.\la{5.30}
\end{align}
Given the transformations (\ref{5.300},\ref{5.30}) we find the transformation of  the momentum.  Its transformation law follows from \eq{5.30} under
condition that \(\int_{\mathbb{R}^4}(\pi \!\cdot\!\mathcal{J})\) is a  scalar
\(\)\begin{align}\la{6.5}
\delta \pi_\alpha&=  \e^\beta(\p_\beta \pi_\alpha -\p_\alpha \pi_\beta)\,+\p_\alpha (\pi_\beta\e^\beta).
\end{align}

The transformation laws  determine admissible variations of the Hamilton principle  (\ref{5.61},\ref{5.71}). 

{ The physical meaning of the admissible variations become clear if we}  treat the diffeomorphism parameter \(\e\)  in \eq{59} as unconstraint D'Alambertian virtual displacements of fluid parcels. Then the flow field \(\mathcal{J}\) and the { canonical momentum $\pi$} change as (\ref{5.30},\ref{6.5}).

\subsection{Equation of motions in the Carter-Lichnerowicz form}\la{4.2}
Let us compute the variation \eq{49} of the action functional, where the variation of the flow field is given by ((\ref{5.30}). After the integration by parts, we obtain 
\begin{align}\la{2.601}
\delta \mathcal{ S}=
-\int_{\mathbb{R}^4}(\mathcal{J}^\alpha\Omega_{\alpha\beta}+\p\cdot\!\mathcal{J\,\pi_\beta})\e^\beta
\,.
\end{align}  
 where the canonical vorticity is defined by \eq{4.200}. Naturally, the same
the result follows from the Hamilton principle in the form \eq{5.71} as \(\delta
\mathcal{ S}=-\delta H=\int_{\mathbb{R}^4}
( \mathcal{J}\!\cdot\delta\pi)\) the admissible variations of the momentum \eq{6.5}.

 The
Hamilton's principle requires the integrand to vanish. Hence,\begin{align}\la{6.8}
\mathcal{J}^\alpha\Omega_{\alpha\beta}+\p\cdot\!\mathcal{J\,\pi_\beta}=0\,.
\end{align}Furthermore, the second term in
\eq{2.601}  vanishes by the
continuity equation \eq{4.4}. 
      The remaining part gives the  Carter-Lichnerowicz equation \eq{1.21} for the
uniformly canonical fluid systems.

\subsection{Conserved currents: helicity and the particle number} 
\subsubsection*{Conservation of the helicity current}{ Hamilton} principle
is the convenient platform to obtain the conservation of the helicity current,
 { a} major property of the uniformly canonical system. That is  \begin{align}
&\p\!\cdot\! h=0\,,\quad h^\alpha=\e^{\alpha\beta\delta\gamma}\pi_\beta\p_\delta
\pi_\gamma\,,\la{1.18}
\end{align}
Let us compute the variation of the action  under a specific diffeomorphism
  \be\e^\alpha=2\eta h^\alpha/(\mathcal{J}\!\cdot\!\pi)\,,\ee where \(\eta\)
is a scalar. { The vector $\e$ }is normal to \(\pi_\alpha\). The
variation  
\eq{2.601} reads
\(\delta \mathcal{_\e S}=
-2\int_{\mathbb{R}^4}(\eta/\mathcal{J}\!\cdot\!\pi)J^\alpha\Omega_{\alpha\beta}h^\beta\).
Next, with the help of identities 
\be 2\Omega_{\alpha\beta}\,{}^\star \Omega^{\beta\gamma}
        =\tfrac 12{ \delta^\gamma_\alpha}\, \Omega_{\delta\beta} \,{}^\star
\Omega^{\beta\delta}=-(\p\!\cdot\! h)\la{6.12}\ee
 followed from \eq{3.121} and the definition of the helicity current \eq{1.18},
  we
obtain 
\begin{align}
\delta \mathcal{_\e S}=\int_{\mathbb{R}^4}\eta(\p\!\cdot\! h)\,.
\end{align} 
This variation must vanish for any \(\eta\). This is possible only if the helicity current is conserved.
\subsubsection*{Continuity equation}
Another specific diffeomorphism \begin{align}
\e^\alpha=-\varphi\mathcal{J}^\alpha/(\mathcal{J}\!\cdot\!\pi)
\end{align} 
where $\varphi$ is a scalar, { entails} the variation  \begin{align}
\delta \mathcal{_\e S}=
\int_{\mathbb{R}^4}\varphi(\p\cdot\!\mathcal{J\,})
\end{align} 
and the continuity equation.  

We observe that in uniformly canonical systems the continuity equation \eq{4.4}  follows from equations of
motion \eq{6.8}.  It does not need to be imposed separately.

\section{Axial current and the axial-current anomaly in Euler fluid}\la{1.2.3}

Given helicity conservation, we obtain the axial-current anomaly. In the
absence of electromagnetic field we identify the  helicity current  with
the axial current \(j_{\sA} ^\alpha=\e^{\alpha\beta\delta\gamma}p_\beta\p_\delta
p_\gamma\).
However,
in the electromagnetic field   helicity current \eq{1.18}  fails to be  a
local
Eulerian
observable as it explicitly depends on the vector potential \(A\). The global
helicity \(\mathcal{H} =\int_{\mathbb{R} ^3}h^0\) is  the gauge invariant,
but its density is not. The problem had been addressed in \cite{abanov2022axial}.
There it was  argued  that  axial current 
 should  be defined as      {\green }\begin{align}
    j_{\sA} ^\alpha=\e^{\alpha\beta\gamma\delta}p_\beta(\p_\gamma
p_\delta+F_{\gamma\delta})\,.
 \la{1.14}
\end{align}
 
Counter to the helicity current \eq{1.18}  the axial current \eq{1.14} is
 a local
Eulerian { field}.  However, it is no longer conserved. The relation \(j_{\sA} ^\alpha=h^\alpha-\e^{\alpha\beta\gamma\delta}\
\p_\beta(A_\delta
p_\gamma)-2{} ^\star\!F^{\alpha\beta}A_\beta\)  and the conservation of the
helicity \eq{1.18} shows that the divergence of the axial current does not vanish, but  is solely expressed through the external electromagnetic field\begin{align}
   \p\!\cdot\!j_{\sA}=\tfrac 12F_{\alpha\beta}\!\cdot\!^\star\!F \,^{\beta\alpha}.\la{1.15}
\end{align}

Eqs.(\ref{1.18},\ref{1.14},\ref{1.15}) represent  the origin of the axial
anomaly: the gauge invariance of observables   conflicts with  the axial gauge invariance. { As the  result the } axial current generated by the axial gauge transformations 
does not conserve.

\section{The flow field and Spacetime covariant Hamiltonian   under the axial coupling}

We return to the axial coupling. Given the explicit form of the action  \eq{3.10}, we may now compute the flow field by exploring the defining relation \eq{49}.  Varying \eq{3.10} over \(\pi\) we obtain (after some algebra) \begin{align}
\delta \mathcal{S}=\int_{\mathbb{R}^4}\pi_\alpha\cdot[\delta n^\alpha+\e^{\alpha\beta\gamma\delta}(\p_\gamma A_\beta^{\sA}\delta p_\delta
+2A^{\sA}_\beta\p_\gamma \delta p_\delta)\,].\nn
\end{align}
This formula defines the flow field \(\mathcal{J}\) up to a term that does not depend on dynamical fields. We chose it such that the difference between \(\mathcal{J}\)  and the electric current \eq{3.40} is a curl
\begin{align}\la{5.2}
\mathcal{J}^\alpha=&n^\alpha-\,{}^\star \! F^{\alpha\beta}_{\sA}p_\beta
+2\e^{\alpha\beta\gamma\delta}A^{\sA}_\beta\p_\gamma \pi_\delta=\nn
\\
&J^\alpha+2\epsilon^{\alpha\beta\gamma\delta} { \p_\gamma
(A^{\sA}_\beta } p_\delta)\,.
\end{align}
 Then the conservation of the electric charge yields the continuity condition \(\p\cdot\mathcal{J}=0\). 

Now, when we know the explicit form of the flow field we obtain the spacetime Hamiltonian by computing the Legendre transform \eq{50}. We obtain a remarkably simple result
 \begin{align}\la{5.5}
{ H}=H^{(0)}-\int_{\mathbb{R} ^4}A^{\sA}\cdot
j_{\sA}.
\end{align}  
In Ref.\cite{abanov2022anomalies} this formula is used as a definition of axial coupling.
Advantage of 
 the functional \(H\) is that it is the explicit function of the momentum and its derivative. This is in contrast to the action  \(\mathcal{S}\)
which, being expressed in terms of \(\mathcal{J}\) does not have a simple form. 

\section{Equations of motion}\la{E}
 \subsection{Euler equation deformed by the axial coupling}
As Eq.\eq{5.2} shows the flow field  \(\mathcal{J}\)
explicitly depends on the axial potential and like the electric current it is not invariant under
the axial-gauge transformation. At the same time,  the equations of
motion \eq{39}  are axial-gauge invariant.  This occurs because  
the last term
in \eq{5.2}   is the null
vector of vorticity tensor as \(^\star\Omega^{\gamma \alpha}\Omega_{\alpha\beta}=0\)
due to helicity conservation \eq{6.12}. It drops from the equations of motions
\eq{1.21}.

The remaining part of \(\mathcal{J}\) \begin{align}
\la{5.17}{I}^\alpha=n^\alpha-{}^\star \! F^{\alpha\beta}_{\sA}p_\beta
\end{align}is
 axial gauge-invariant and so the equations of motions. Also, \(I^\alpha
\) does not depend on the electromagnetic field. 

In terms of $I^\alpha$ the Carter-Lichnerowicz equation   \eq{1.21} and  the continuity
equation \eq{4.4} read \begin{align}\begin{cases}
{I}^\alpha\Omega_{\beta\alpha}=0\,,\la{5.7}\\
\p\!\cdot\! {I}=\,{}^\star\!F_{\sA}\cdot\Omega\,.
\end{cases}
\end{align}
We may bring these equations  the form close to
Eq.\eq{29}. Introducing \be{\bm V}:=\frac{\bm
I}{I^0}=\frac{\bm n-\bm p\!\times\!  \bm E^{\sA}\,+p_0\, \bm B^{\sA}}{n -{\bm
B}^{\sA}\!\cdot\!
\bm p}\ee and after some algebra, we obtain
\begin{align}\la{3.6}\begin{cases}
\dot{\bm \pi}-\bm \nabla \pi_0-\bm V\!\times \!(\bm\nabla\!\times
\!\bm \pi)=0\,,\\
\dot n+ \bm\nabla\!\cdot\bm n=2 (\bm E^{\sA}+ \bm V\!\times\!\bm B^{\sA}
)\!\cdot (\bm\nabla\!\times\!
\bm \pi)\,.
\end{cases}
\end{align}
This form emphasizes the effect of the axial field as a deformation
of the Euler equation and the continuity equation of the perfect fluid.
{ We observe that $\int_{\mathbb{R}^3} n$ is no longer an integral. The fluid exchange their particles with the reservoir. }

\subsection{Direct derivation of equations of motion in the form of
 conservation laws }\la{B}

Although it may not be obvious, the equations \eq{5.7} are equivalent to
Eq.\eq{39}. A direct check requires a good deal of uninspiring algebra.
 More instructive is to obtain the equations of motion  \eq{39} directly from the Hamilton principle \eq{5.61}. { This  will be a prove that the action has the form \eq{6}  and is consistent with the (\ref{341},\ref{3.2}) despite of apparent sign difference of the axial coupling.
 
We} compute the variation of the action \eq{6} (repeated below)
\begin{align}\la{4.100}
\mathcal{S}=\mathcal{S}^{(0)}+\int_{\mathbb{R} ^4} \left[A\cdot J-A^{\sA}\cdot
j_{\sA} \right]\,
\end{align}  
{ under  variations of vector fields \eq{5.30}}.

We start from the  action \(\mathcal{S}^{(0)}\).  Its variation follows from
\eq{2.601} under the  identification the particle number current \(\mathcal{J^\alpha}=n^\alpha\)
and \(\Omega_{\alpha\beta}=\p_\alpha p_\beta-\p_\beta p_\alpha\).     We
obtain the divergence of the momentum-stress-energy tensor
plus the divergence of \(n^\alpha\) \be\delta \mathcal{S}^{(0)}=\int_{\mathbb{R}
^4}  [-\p_\alpha T^\alpha_\beta
+p_\beta\p_\alpha n^\alpha]\e^\beta \,.\ee 
{ It is instructive to notice a difference between this result  and  the change of the action under coordinate transformation  \eq{341}. The difference vanishes in the perfect fluid where}  \(\p_\alpha n^\alpha=0\). With the help of the continuity equation
\eq{4.4}, the explicit form of the { flow field} \eq{5.2} and
the identity \eq{6.12}
we obtain the relation\be p_\beta\p_\alpha n ^\alpha=
2 F_{\beta\alpha}^{\sA}j^\alpha_{\sA}\,.\ee
Therefore,\begin{align}
\delta \mathcal{S}^{(0)}=\int_{\mathbb{R} ^4}  [-\p_\alpha T^\alpha_\beta+2
F_{\beta\alpha}^{\sA}j^\alpha_{\sA}]\e^\beta
\,.\la{7.4}
\end{align}
The next step is to find the variation of the couplings in \eq{4.100}. We
get 
\begin{align} \nn
\delta\Big(\int_{\mathbb{R}^4} A\cdot J \Big)=\int_{\mathbb{R}
^4}\e^\beta\Big[F_{\beta\alpha}J^\alpha+A_\beta (\p\!\cdot
\!J)\Big] =
&\\ 
\nn
\int_{\mathbb{R} ^4}\e^\beta F_{\alpha\beta}J^\alpha=
\int_{\mathbb{R} ^4} \e^\beta \Big[F_{\beta\alpha} j^\alpha+2F_{\beta\alpha}
\,{}^\star\!
F^{\alpha\gamma}A^{\sA}_\gamma\Big]
&\\ =
\int_{\mathbb{R} ^4} \e^\beta \Big[F_{\beta\alpha}
j^\alpha+\tfrac 12 A^{\sA}_\beta F_{\gamma\alpha} \,{}^\star\!
F^{\alpha\gamma}\Big]\nn=
&\\
\int_{\mathbb{R} ^4} \e^\beta \left[F_{\beta\alpha}
j^\alpha+A^{\sA}_\beta
(\p\!\cdot\!
j_{\sA})\right],\la{5.6}
\end{align}
\begin{align}
-\delta\left(\int_{\mathbb{R} ^4} A^{\sA}\cdot j_{\sA} \right)=&-
\int_{\mathbb{R} ^4}\e^\beta\left[F^{\sA}_{\beta\alpha}j_{\sA}^\alpha+A^{\sA}_\beta
(\p\!\cdot\!
j_{\sA})\right]\la{7.6}
\end{align}
In  these equations, we took into account that    the 
currents are transformed according to the rule \eq{5.30},  that the electric
current is divergence-free,  the relation between  the  electric 
current and  the vector current  \eq{2.300}, the identity \(2F_{\beta\alpha}
\,{}^\star\!
F^{\alpha\lambda}=\tfrac 12\delta_\beta^ \lambda F_{\gamma\alpha} \,{}^\star\!
F^{\alpha\gamma}\)   and
the divergence of the axial
current
\eq{1.15}.  Combining  the contributions of (\ref{7.4}-\ref{7.6})
 we notice that the last terms in \eq{5.6} and \eq{7.6} cancel each other
and the last term in \eq{7.4} changes the sign in the first term of \eq{7.6}.
 The result  is  
\begin{align}
\delta \mathcal{S}=\int_{\mathbb{R} ^4}\e^\beta\left[ -\p_\alpha T^\alpha_{\beta}+
F_{\beta\alpha} j^\alpha  + F^{\sA}_{\beta\alpha} j^\alpha_{\sA} \right]\,.
\end{align}
That prompts Eq.\eq{39} and explains the choice of an unusual sign in the
axial coupling in \eq{4.100}. 
\section{Discussion}\la{D}
We highlight the major points of the paper.

We developed the spacetime covariant Hamilton principle on the Lie algebra
$\rm{Diff}(\mathbb{R}^4$) and used it to study the effects of
the chiral anomaly in hydrodynamics. In particular, we obtain the set of conservation slaw which give the full set of equations of motion. For references we collect them here 
 \begin{align}\la{79}\begin{cases}
\p_\alpha T^\alpha_\beta &= F_{\beta\alpha}j^\alpha+F^{\sA}_{\beta\alpha}j_A^\alpha\,,\\
\p_\alpha
j^\alpha_{\sA}&=    \tfrac  12{}^\star\!F^{\alpha\beta}\!\, F_{\beta\alpha}\,,\\
\p_\alpha j^\alpha &= F_{\alpha\beta}\, {}^\star\! F^{\beta\alpha}_{\sA}\,.
\end{cases}\end{align}  
Also, we found the hydrodynamic representations of the vector and the axial current. We collect them here 
\begin{align}\la{80}
\begin{cases}
j^\alpha=n u^\alpha+^\star \!\! F_{\sA}^{\alpha\beta}p_\beta\,\,,\\
j_{\sA} ^\alpha=\e^{\alpha\beta\delta\gamma}p_\beta(\p_\delta
p_\gamma+F_{\delta\gamma})\,.
\end{cases}
\end{align} 
Given the explicit form of the stress tensor and the currents, the conservation laws provide six equations for four Eulerian fields \(n\) and \(\bm v\). Various consistency conditions we used in the text warrant that this overdetermine set is consistent. Equation \eq{80} could be seen as a realization of the overcomplete set of conservation laws (79).

We identified the class of fluids where the chiral anomaly
takes place. These are uniformly canonical fluid systems.
Such systems could be defined in different ways. One of
them is to say that spacetime momentum is the only
dynamic variable of the variational functional. This natural
property gives rise to the foliation of the spacetime by
vorticity surfaces, the geometric root of the conservation of
helicity, and the chiral anomaly.

The spacetime covariant Hamilton principle is commonly used in relativistic
fluid dynamics.
It could be used for non-relativistic hydrodynamics but often does not bring
{ a} new content if the action of the Lie derivative \(\mathcal{L}_\e\) which
evaluates a change of the  spacetime tensors along a spacetime \(\mathbb{R}^4\) vector
field \(\e\) is essentially reduced to the action of \(\p_t+\mathcal{L}_{\bm\e}\),
where \(\bm \e\) is the spatial \(\mathbb{R}^3\) vector field as it happens
in the Galilean perfect fluid.  This is no longer true if there is the axial
``magnetic''  \(\bm B^{\sA}\).  In this case, the spacetime covariant approach
used in the paper is indispensable.

In the nutshell the chiral anomaly could be formulated as follows: equations
of motions, such as \eq{39}  are composed of the fields tensors of vector
and axial potentials \(F_{\alpha\beta}\) and \(F_{\alpha\beta}^{\sA}\). They
are gauge invariant under vector and axial gauge transformations (\ref{1.8},\ref{1.9})  evaluated at given Eulerian fields. 
 Solutions to these equations being invariant under vector gauge transformation
\eq{1.8}  may not be necessarily invariant under the axial transformation
\eq{1.9}. This is also true for physical observables such as electric current
\eq{3.21}. Important solutions, such as the stationary
flow explicitly depend on the value of the axial potential.
In  \cite{wiegmann2022chiral} it had been shown that the stationary solutions
are Beltrami flows.  Let us recall the major result of  \cite{wiegmann2022chiral}.  To simplify the matter, let's turn the electromagnetic
field off. Then if the remaining axial  background potential is time-independent,
the stationary flow exists and is given  by the condition \(p_0=0\)  and
the Beltrami condition, that is  the ``velocity"  \(\bm V\) and vorticity
\(\bm\nabla\!\times\!\bm p\) are collinear \cite{arnold2008topological}.
The Beltrami condition satisfies the first equation \eq{3.6}. Then the second
equation \eq{3.6} reduces to the equation \(\bm\nabla\cdot \bm n=2 \bm E^{\sA}\!\cdot
(\bm\nabla\!\times\!
\bm p)\) whose solution is Beltrami flow

\be
    {\rm curl}\,\bm { v}=\kappa\bm {\bm v}\,,
\ee 
where \({\rm\ curl}\, \bm v=\tfrac 1{\sqrt{|A_0^{\sA}|}}\bm\nabla(\sqrt{|A_0^{\sA}|}\times\bm
v)\)
and the  Beltrami coefficient \(\kappa=n/2mA_0^{\sA}\)\, is determined by
the axial-vector potential.

\section*{Acknowledgement}
The work of P.W. was supported
by the NSF under Grant NSF DMR-1949963. The author acknowledges collaboration
with   A.G. Abanov on this project. Also, the author thanks for the hospitality
and support of Galileo Galilei
Institute during  the  program on    "Randomness, Integrability, and Universality"
  and the International Institute of Physics in Natal where this work had
been completed
and  the support of  the Simons Foundation program for
GGI Visiting Scientists. 


\appendix
\section*{ Spacetime covariant Hamilton principle and   conservation laws  in terms of differential forms} 
In this Appendix, we recast the equations of motions and conservation laws of the uniformly canonical fluid systems in terms of differential forms emphasizing some geometric aspects of the flows. 

We describe the fluid configuration space by the vector field $\mathcal{U}$   dual to the current 3-form $\mathcal{J}$.  The formal relation is: the current 3-form  is the volume form evaluated on the vector field $\iota_{\mathcal{U}}{\rm vol}=\mathcal{J}$. We also consider the momentum  1-form $\pi=\pi_\alpha dx^\alpha$.

Under a diffeomorphism, $x^\alpha\to x^\alpha+\e^\alpha$ the change of a tensor defines the Lie derivative $\mathcal{L}_\e$ along the vector field $\e^\alpha\p_\alpha$. The action of the Lie derivative   on a differential form is expressed   in terms of the interior
product and the exterior
derivative by the Cartan formula
  \be\mathcal{L}_{\e}=\iota_\e d+d \iota_\e\,.\la{2.8} \ee
  We recall that the interior product \(\iota_\e\) acting on a differential
form is the mere contraction of the cotensor with the vector  \(\e\). 

Applying the the Cartan formula we obtain  Lie derivative to the current 3-form $\delta\mathcal{J}=\mathcal{L}_{\e}\mathcal{J}$ and the momentum 1-form $\delta\pi=\mathcal{L}_{\e}\pi$. In components they are given by (\ref{5.30},\ref{6.5}). 
 
Using these formulas we obtain the variation of the action. Say, we  choose the natural variables to be \(\pi\). Then 
\begin{align}
\delta\mathcal{S}
 =-\int_{\mathbb{R}^4}
\iota_\mathcal{U}\delta\pi=\int_{\mathbb{R}^4}\iota_\e\iota_\mathcal{U}
\Omega \,,\la{4.20}
\end{align}  where we used the rule of the interior product
 \(\iota_\mathcal{U}\iota_\e=-\iota_\e\iota_\mathcal{U}\). Then the Carter-Lichnerowicz
equation   follows \cite{markakis}
\begin{align}\la{87}
\iota_{\mathcal{U}}\Omega=0\,.
\end{align}
\paragraph*{Lagrange-Cauchy form of Euler equation and Helmholtz law.} 
Let us examine the
advection of the fluid momentum. That is the Lie derivative of the momentum
along
the flow vector field  \(\mathcal{U}\).   Applying the Cartan
 formula  \eq{2.8} we write \(
\mathcal{L}_\mathcal{U}\pi=\iota_\mathcal{U}\Omega+d(\iota_\mathcal{U}\pi)
\). 
Then by the equation of motion  \eq{87}, we obtain the Lagrange-Cauchy
form of the Euler equation\begin{align}\la{89}
\mathcal{L}_\mathcal{U}\pi=d(\iota_\mathcal{U}\pi).
\end{align}
 The Eq.\eq{89}  represents the fluid dynamics as a degree
to which the acceleration deviates from
 streamlines. 

Taking the exterior derivative of \eq{89} and using its commutativity with the Lie derivative we obtain the Helmholtz law:  canonical vorticity
2-form \(\Omega=d\pi\) is a frozen-in substance advected by the flow along
the fluid streamlines \begin{align}
\mathcal{L}_\mathcal{U}\Omega=0\,.\la{2.16}
\end{align}  
As a consequence, the vorticity flux
through an area encompassed by a fluid
contour co-moving with the flow does not change in time (Kelvin theorem).

\paragraph*{Foliation}
{  The Carter-Lichnerowicz equation states that the vorticity matrix is rank-2 degenerate having  two zero-eigenvectors $\mathcal{J}^\alpha$ (streamlines)  and $^\star\Omega^{\alpha\beta}\mathcal{J}_\beta$.  This entails a special geometric property of uniformly canonical systems 
 \cite{carter}: the spacetime is foliated by a family of integral vorticity 2-surfaces  whose tangent vectors are are null vectors of vorticity.
}
In other words, streamlines define a family of smooth 2-dimensional vorticity surfaces each passing through a given point of spacetime.

  \paragraph*{Conservation of helicity } 
The conservation of helicity \eq{1.18} could be traced  to the  foliation property with no reference to  a relation between the flow field  \(\mathcal{J}\)
and the fluid momentum. 

The  divergence of the helicity current  is the exterior derivative \(dh\) of the helicity 3-form \(h=\pi\wedge d\pi\). It is the top-form
 \begin{align}
dh=\Omega\wedge \Omega\,.
\end{align}
Using the rule for the interior product \be \iota _{X}(m \wedge l
)=\left(\iota _{X} m\right)\wedge l +(-1)^{q}m \wedge \left(\iota _{X} l\right)\la{2.11}\ee
where \(l\,, m\) are differential forms and \(q\) is the degree of the form
\(m\), we obtain \(\iota_\mathcal{U}(dh)=\iota_\mathcal{U}(\Omega\wedge
\Omega)=2\Omega\wedge (\iota_\mathcal{U}\Omega) \) which vanishes due to \eq{87}. 
 Hence,
  \(dh=0\). 

{ Alternatively, we notice that \(dh=\Omega\wedge \Omega={\rm Pf}[\Omega_{\alpha\beta}]\)
, where ${\rm Pf}=\sqrt{{\rm det}[\Omega_{\alpha\beta}]}$ is the Pffafian of the $4\times 4$ matrix $\Omega_{\alpha\beta}$. Since this matrix has  null vectors  its determinant and the Pffafian vanish.
}

 \bibliography{helicity2}

\end{document}